\documentclass[a4paper,fleqn,usenatbib]{mnras}

\usepackage{newtxtext,newtxmath}

\usepackage[T1]{fontenc}
\usepackage{ae,aecompl}

\usepackage{graphicx}	
\usepackage{amsmath}	
\usepackage{amssymb,latexsym}	
\usepackage{color}

\renewcommand{\vec}[1]{\boldsymbol{#1}}

\newcommand{\cm}{\ensuremath{\,{\rm cm}}}

\newcommand{\pc}{\ensuremath{\,{\rm pc}}}

\title[Radio burst detection by Hough transform]{A Radio Burst Detection Method Based on the Hough Transform}

\author[Zuo \& Chen]{
Shifan Zuo,$^{1,2,3}$
Xuelei Chen,$^{1,2,4}$\thanks{E-mail: xuelei@cosmology.bao.ac.cn}
\\
$^{1}$Key Laboratory for Computational Astrophysics, National Astronomical Observatories, Chinese Academy of Sciences, Beijing 100101, China\\
$^{2}$School of Astronomy and Space Science, University of Chinese Academy of Sciences, Beijing 100049, China\\
$^{3}$Department of Astronomy and Tsinghua Center for Astrophysics, Tsinghua University, Beijing 100084, China\\
$^{4}$Center of High Energy Physics, Peking University, Beijing 100871, China
}

\date{Accepted XXX. Received YYY; in original form ZZZ}

\pubyear{2019}

\begin{document}
\label{firstpage}
\pagerange{\pageref{firstpage}--\pageref{lastpage}}
\maketitle

\begin{abstract}
  We present a simple and fast method for incoherent dedispersion and fast
  radio burst (FRB) detection based on the Hough transform, which is  widely used for feature 
  extraction in image analysis. The Hough transform maps a point in
  the time-frequency data to a straight line in the
  parameter space, and points on the same dispersed $f^{-2}$
  curve to a bundle of lines all crossing at the same point, thus the curve is
  transformed to a single point in the parameter space, enabling an easier way for the detection 
  of radio burst. By choosing an appropriate truncation threshold, in a reasonably radio quiet environment, 
  i.e. with radio frequency interferences (RFIs) present but not dominant, the computing speed of the method is very fast.
  Using simulation data of different noise levels, we studied how the detected 
  peak varies with different truncation thresholds. We also tested the method with some 
  real pulsar and FRB data.
\end{abstract}

\begin{keywords}
 radio continuum: transients, methods: data analysis
 \end{keywords}



\section{Introduction}\label{S:intro}
Astronomical radio pulses are dispersed while traveling through the
interstellar medium (ISM) or intergalactic medium (IGM) plasma. 
At a lower frequency the wave travels at a lower speed  and arrives at a later time. 
This dispersion of the arrival time significantly decreases the pulse 
amplitude at a fixed observation time. In order to improve the detection sensitivity, 
dedispersion of the signal is required to compensate for the time
delay induced by dispersion. A number of dedispersion and detection 
algorithms have been developed over the years, 
the computation is demanding as it often needs to be done in nearly real time. 
For the recently discovered fast radio bursts (FRBs), which are bright millisecond 
radio pulses with unknown origin and mostly non-repeating, this is especially so. 
The inferred FRB rate is fairly high \citep{Lorimer2007,Thornton2013,Petroff2015,Amiri2017}. 
Efficient dedispersion algorithms would be very useful for searching FRBs.

The received signal can be de-dispersed by applying frequency dependent time
delays to the signal prior to integration, but the difficulty is that usually 
the amount of dispersion is not known, so a large number of trials with different
dispersion measures have to be attempted in each search. This brute force dedispersion procedure requires expensive computations,
of a complexity $O(N_{t} N_{f} N_{d})$, where $N_{f}$, $N_{t}$ and $N_{d}$ are the dimension 
of the data in  frequency, time, and dispersion measure, respectively.
To speed up the dedispersion process, many algorithms have been
developed, for example, the tree dedispersion algorithm, which has a
complexity of $O(N_{t} N_{f} \log N_{f})$ \citep{Taylor1974}, the Fast
Dispersion Measure Transform (FDMT) algorithm of complexity $O(2
N_{t} N_{f} + N_{t} N_{d} \log_{2}N_{f})$ \citep{Zackay2014},
etc. 

Obviously, a sensitive dedispersion algorithm should maximise the
signal-to-noise ratio of the pulse, which can only be fulfilled by
integrating the flux exactly along  the dispersion curve in the
time-frequency domain. Mathematically, detection of such a curve 
can be achieved  by a family of transformations, for example, the 
Radon transform \citep{Radon1917} and Hough transform \citep{Hough1962}. 
The Radon transform maps a curve or more generally a shape 
$c(\vec{p})$ in a $D$-dimensional space to a parameter space by integral projection, 
\begin{equation} \label{eq:radon}
  \mathcal{R}_{c(\vec{p})}\{I\} = \int_{\vec{x} 
    \in c(\vec{p})} I(\vec{x}) \, dc  
    = \int_{\mathbb{R}^{D}}
  I(\vec{x}) \, \delta(\mathcal{C}(\vec{x}; \vec{p})) \, d\vec{x},
\end{equation}
where $\vec{p}$ is a vector of parameters describing the shape,
$\mathcal{C}(\vec{x}; \vec{p})$ is a set of constraint functions
that together define the shape, and $\delta(\cdot)$ denotes the Dirac delta
function in the above, or the Kronecker delta in the discrete case. 
The Hough transform is closely related to the Radon transform \citep{Ginkel2004}, 
though in its original formulation it is inherently discrete. 
It was originally designed to detect straight lines in binary
images, but it can be extended to detect more general
shapes and in grey-valued images. For this purpose, we set up an $N$-dimensional accumulator
array $A(\vec{p})$, each dimension of it corresponding to one of the
parameters of the shape to be searched. Each element of this array
contains the number of ``votes" in favour of the presence of a shape with the
parameters corresponding to that element. The votes are
obtained as follows: for each point $\vec{x}_{i}$ with value $g_{i} = I(\vec{x}_{i})$
in the input image $I(\vec{x})$, if the shape passes through it,  the 
vote for this shape parameter is increased by an amount of $g_{i}$, i.e., let 
\begin{equation}
A(\vec{p}) \leftarrow A(\vec{p}) + g_{i} \, \delta(\mathcal{C}(\vec{x}_{i}; \vec{p})).
\end{equation}
If a shape with parameter $\vec{p}$ is present in the image, all of the pixels
that are part of it will vote for it, yielding a large peak in the
accumulator array. The shape detection problem in the image space is then transformed to a
simple peak finding problem in the parameter space. 
As we usually do not know the dispersion measure in advance, the whole parameter
space (in practice a range of dispersion measures) needs to be
explored. Using the fact that most points in the data are background noise, 
we could truncate the data according to an appropriate threshold, this will
throw away most of the noise below the threshold, thus making the map sparse, and
the required computation is then drastically reduced. 

The use of Hough transform 
for radio transients detection and dedispersion was investigated in
\cite{Fridman2010}, in which a dispersed pulse is approximated as a straight line 
in the time-frequency plane within a small bandwidth. The data is first converted to a binary image, 
by taking a threshold  given by $1 \sigma$ value above the mean. The method was demonstrated with 
the application of Hough transform to the pulsar B0329+54 data observed 
by LOFAR in 10 MHz bandwidth. 

In this paper we study the detection of radio pulses with the Hough transform. 
We do not make the  straight line approximation
but detect directly  the  $f^{-2}$ pulse track curve on the time-frequency plane,
hence not limited in the usable bandwidth. In the truncation we will not fix the threshold, 
but use robust statistical quantities based on
the median and median absolute deviation (MAD) to determine the
threshold value, which are more reliable and less affected by
outliers and strong pulse signals presented in the
data. We apply the Hough transform to the
truncated gray-valued image instead of the binary image, to help 
suppress the noise and improve the signal-to-noise ratio in 
the transformed parameter space. 

\section{Algorithm}
\label{S:algorithm}
We consider the Hough transformation algorithm for incoherent dedispersion and pulse search. 
 Our input data is a time stream of spectrum which is the short time integral of the intensity, 
either from a single receiver or from the synthesised beam of an array.  
The dispersion delay of the pulse arrival time at a frequency
$f_{1}$ relative to $f_{2}$ is given by 
\begin{equation} \label{eq:dt1}
  \Delta t = t_{1} - t_{2} = d \, (f_{1}^{-2} - f_{2}^{-2}).
\end{equation}
where $t_i$ is the arrival time of signal at frequency $f_{i}$ in units of ms, 
$d \equiv \frac{e^{2}}{2\pi m_{e} c} \cdot \text{DM}
  \simeq 4.15 \times \text{DM}$, where $e$ is the
  elementary charge, $m_{e}$ is the electron mass, $c$ is the speed
  of light in vacuum, and DM  is the dispersion measure
in units of $\pc \cm^{-3}$, $f_{i}$ is frequencies measured in GHz.
Each dispersed pulse signal arrival time at frequency $f$ then falls on a curve
\begin{equation} \label{eq:tc}
  t = d \, f^{-2} + t_{0},
\end{equation}
where $t_{0}$ is the time offset of the curve, which can be uniquely determined by the two parameters $(d, t_{0})$.

\subsection{Hough Transform}
\label{S:hough}

For a data point $(t_{i}, f_{i})$ on the curve defined by Eq.~(\ref{eq:tc}), we
have the relation of the two parameters as
\begin{equation} \label{eq:t0d}
  t_{0} = -f_{i}^{-2} d + t_{i},
\end{equation}
which in the parameter space $(t_0, d)$ is a straight line with slope $-f_{i}^{-2}$ and
interception $t_{i}$, so each point on the curve defined in Eq.~(\ref{eq:tc}) in the data 
maps to a straight line in this space, and all of the points on the curve map to a bundle of lines,
which all cross at the same point $(t_{0}, d)$. The problem of detecting a $f^{-2}$
curve in the observing data is transformed to a peak detection problem, 
which is both easier and more robust.
Furthermore,  the presence of discontinuity and outliers have
little effect on the peak detection in the
parameter space, as long as there are enough identifiable points on
the curve. Outliers, even ones in the form of a line, will not generate 
peaks as high as the one corresponding to the curve since they do not
have the $f^{-2}$ function form.

To apply the Hough transform, we
initialise an all zero accumulator matrix $A(t_{0}, d)$ of dimensionality $N_{t_{0}} \times N_{d}$,
with DM range $[d_{\text{min}}, d_{\text{max}}]$. For each
point $(t_{i}, f_{i})$ in $I$, we accumulate a straight line given by 
Eq.~(\ref{eq:t0d}) with strength $I(t_{i}, f_{i})$ to the accumulator
$A$, i.e., 
\begin{equation}
A \leftarrow A + I(t_{i}, f_{i}) \, \delta(f_{i}^{-2} d -t_{i} + t_{0}).
\label{eq:Aa}
\end{equation}
This will take $O(N_{d})$ operations. We see lines
corresponding to points that are on the pulse curve Eq.~(\ref{eq:t0d})
will all cross at the point $(t_{0}, d)$, generating
a high peak at this point in $A$, with value about $\mu N_{s} $ where $N_{s}$ is
the number of points on the curve and $\mu$ is
the mean value of these points. Because the accumulation
  operation is order-independent, the Hough transform can be
  naturally parallelised by partitioning the data points in $I$, this is
  true for the background truncated image we will discuss later, too.

If the source dispersion value $d_{s}$ is known \textit{a prior}, as in the case of known pulsars, 
the DM range can be very narrow, otherwise a wide range should be chosen to cover 
possible dispersion for the searched signal.  
Once we have chosen the appropriate range $[d_{\text{min}}, d_{\text{max}}]$, 
the range of $t_{0}$ is
\begin{eqnarray}
t_{0,\text{min}} &=& -d_{\text{max}} f_{\text{min}}^{-2} +t_{\text{min}},\\
t_{0,\text{max}} &=& -d_{\text{min}} f_{\text{max}}^{-2} +t_{\text{max}},
\end{eqnarray} 
and $N_{t_{0}}\approx N_{d}$.
The attainable resolution of $d$ is determined by the 
time and frequency resolution: from Eq.~(\ref{eq:dt1}), for neighbouring frequency
$d \approx \frac{1}{2} \Delta t f^{3} / \Delta f,$ 
so
\begin{equation}
\Delta d = \frac{1}{2} \Delta t \frac{3 f^{2} \Delta f}{ \Delta f} = \frac{3}{2} f^{2} \Delta t \sim
f_{\text{min}}^{2} \Delta t.
\end{equation}
Conversely, given a maximum size of the data that could be stored,
the maximum time duration then limits the range of dispersion to be searched in full efficiency.

In practice, the dedispersion and pulse search is done within a data frame of finite time length. It is quite 
frequent that the dispersed pulse signal lasts beyond the data frame in which it was initiated. To avoid losing sensitivity to 
such signal, the data frame must partially overlap with each other, such that the dispersed signal can be captured fully within 
one data frame. The dedispersion and pulse search algorithm is applied to successive partially overlapping time stream data. 
Below we shall compare the Hough transform to other algorithms on a single frame basis, but note that the Hough transform 
algorithm is well suited for such sliding window processing, as we can cache
and reuse the accumulator array of the overlapping area to further reduce the computations. For example, 
if the overlap area is half of the data frame, we can compute the Hough accumulator $A_{i1}$ and $A_{i2}$ for the two halves of
the $i$th frame separately, then the total is $A_i=A_{i,1}+A_{i,2}$, but then for the next frame, we will have $A_{i+1,1}=A_{i,2}$.

\subsection{Background Subtraction and Thresholding}
\label{S:subtraction}

Before applying the Hough transform, we first pre-process the data by subtracting out the mean of the 
background in the time-frequency data frame. We can apply the Hough transform to 
this background-subtracted data, but then the computation is
inefficient, as most of the data are just noise, while the pulse signal if
present only takes a very small portion of the data. 
In this case, every point in the data is to be
transformed, leading to a worst case computational complexity of
$O(N_{t} N_{f} N_{d})$ which is the same as that of the brute force
dedispersion procedure.  To reduce the amount of computation, we can apply a truncation threshold to 
filter out most of the data before doing the Hough 
transformation. We can record in a data structure such as an one dimensional array or a stack
the array indices or memory pointers of the data points $(t_{i}, f_{i})$ 
whose background-subtracted value $|I(t_{i}, f_{i})|$ is 
higher than the truncation threshold, this flattens the sparse background
  filtered 2d array into a 1d array of much fewer data points. We then do Hough transform according to
Eq.~(\ref{eq:Aa}) for only these data points by searching within
this array. 
For each such point $(t_{i}, f_{i})$, in the accumulator array $A$ we add a value $I(t_{i}, f_{i})$ 
to all data points $(d_{j}, t_{0,j})$ for $j = 1, \cdots, N_{d}$ located on the straight 
line Eq.~(\ref{eq:t0d}). For each $d_{j}$ the corresponding $t_{0,j} = -f_{i}^{-2} d_{j} + t_{i}$.

After the thresholding, if the number of non-zero pixels in the now sparse image is $N$, and
for each point it takes $O(N_{d})$ Hough accumulation operations, then
the computation has a complexity of order $O(N N_d)$.  
To assess $N$, first consider the ideal case with no radio frequency interferences (RFIs). 
If the truncation threshold is set appropriately, such that most of the
pulse is preserved while the pixels with only noise are zeroed, 
then the number of points on the curve for a pulse with maximum length 
is $\sim \text{max}(N_{t}, N_{f})$. So when the pulse signal is present, 
 $N \sim O(\text{max}(N_{t}, N_{f}))$. However, in most data frames the pulse signal will not be present, so 
 on average $N \ll O(\text{max}(N_{t}, N_{f}))$. Note however the computational complexity is significantly 
 reduced only for relatively high threshold values ($\tau \gtrsim 3$).
  
However, in the real world the RFIs are generally present, and the RFI-contaminated
 pixels would have large values that allow them to pass the thresholding, so $N$ would be determined by the 
 number of such pixels. In the typical radio astronomy cases, the RFIs are present but not dominant, i.e. only 
 a small fraction of particular frequency and time bins are affected, then again $N  \sim O(\text{max}(N_{t}, N_{f}))$.
 So in the end the computation complexity is $O(\text{max}(N_{t}, N_{f}) N_d)$. Here the
  computational complexity does not include the cost of background subtraction and thresholding, which 
 is an operation of complex order $O(N_t N_f)$ and takes negligible amount of computing than the 
  more complicated accumulation for multiple dispersions. Note this estimate of complexity is valid only 
  in a very limited scope, it would fail if the RFIs are not narrowly distributed in the time-frequency domain as we assumed, 
  or if the fraction of the data contaminated by RFIs is significant.

For a  background noise with a Gaussian distribution $N(\mu,
\sigma^{2})$, which is a good approximation for receiver noise or
astronomy background in a short period of time, the threshold can be
set as $T = \tau \sigma$, which will remove $\sim 68\%$, $\sim 95
\%$ and $\sim 99.7 \%$ of the data for $\tau = 1.0$, $2.0$ and $3.0$
respectively. In the truncation process, some pulse signal may also
be thrown away, especially for low signal-to-noise ratio (SNR)
data. The threshold $T$ should be chosen to achieve good sensitivity
while reducing the computing time to a practical level.

If strong outliers such as radio frequency interferences (RFIs) are present in the data, 
the Gaussian model of noise may not be valid. The more robust median and the median absolute
deviation (MAD) may be used instead \citep{Hampel1974,Rousseeuw1993,Leys2013,Fridman2008}. We set 
\begin{eqnarray}
\hat{\mu} &=& \text{median}(I), \\
\hat{\sigma} &=& \text{MAD}(I) \equiv \text{median}(|I -\text{median}(I)|) / 0.6745.
\end{eqnarray}
 In practice,  the median and MAD do not need to be computed every cycle, but can instead be 
 updated after a number of cycles to reduce the amount of computation. 
 There is a small chance that along one of the curves the data points happen to fluctuate in such a way 
that  they generate a peak in the accumulator matrix. However,  the background 
mean has already been subtracted, the remaining noise has a zero mean truncated Gaussian 
distribution, so their expectation value would be 0, as the positive and negative values of the 
noise will typically cancel out in the
accumulation.

\subsection{Outliers}\label{S:disc}

Outliers such as RFIs are often present in the data,  which may be much stronger than the astronomical radio
pulse signal. In most cases, however, the outliers would appear as vertical (short pulse in time) or 
horizontal (narrow frequency band) lines in the data $I(t, f)$.  
Such outliers are automatically filtered out,  as these data points will be mapped into lines which will 
cross at $d=0$ (no time delay) or $d=\infty$ (infinitely large dispersion), which would not contribute to the 
accumulator matrix in the reasonable range of $[d_{\rm min}, d_{\rm max}]$.
It is not very likely that the shape of the outliers happens to 
appear as a $f^{-2}$ curve with parameter $(d, t_{0})$ in
the right range, though in rare coincidence such event could be produced, e.g. as
in the case of the so called ``peryton'' which has a roughly $f^{-2}$ shape \citep{Petroff2015b}. A simple automated algorithm 
as discussed here may not be able to identify all such cases, but hopefully the algorithm could 
filter out most outliers such that only a small number of pulse events remain and can be 
further investigated in detail with human intervention.

\begin{figure}
  \centering
 \includegraphics[trim=10 10 40 34,clip,width=0.38\textwidth]{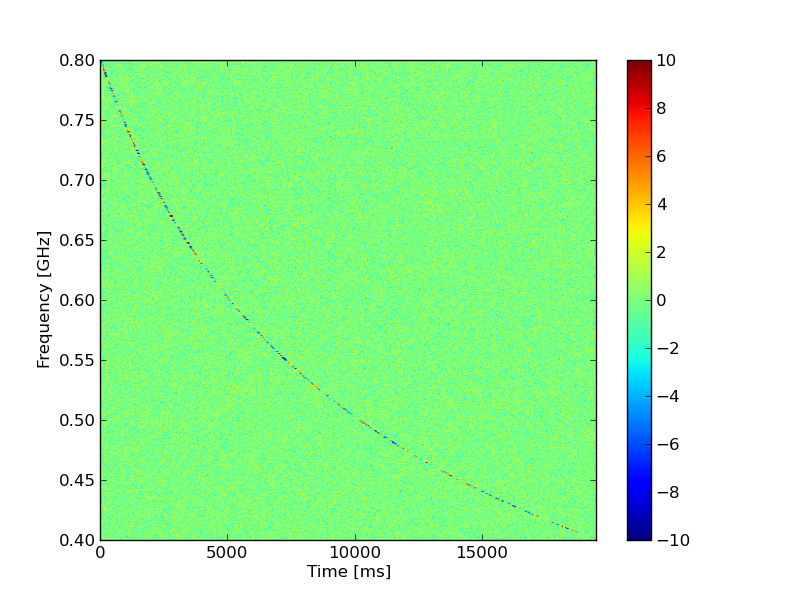}
 \includegraphics[trim=10 10 40 34,clip,width=0.38\textwidth]{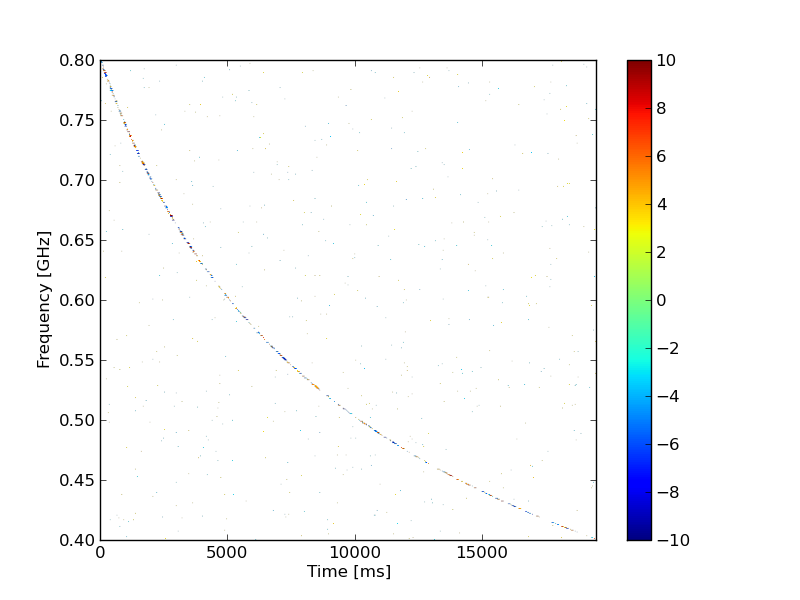}
  \caption{Top: simulated data; Bottom: truncated data (with $\tau = 3.0$)
    which masks out most of the noises. }
  \label{fig:sim}
\end{figure}


\section{Application Test}
In this section we test how our algorithm works with data. We first test it with
simulation data (\ref{S:sim}), then test with real pulsar and FRB data (\ref{S:rd}).

\subsection{Simulation Data}\label{S:sim}
 We generate a mock sample of observing data as a 
 superposition of Gaussian noise and a dispersed FRB signal.
The noise is independent and identically distributed (iid) Gaussian
with a distribution $N(0, \sigma_{n}^{2})$, and the signal is also
iid Gaussian with distribution $N(\mu, \sigma_{s}^{2})$. The
parameters for the simulated data are set as $\sigma_{n} = 1.0$, $\mu
= 3.0$, $\sigma_{s} = 3.0$, and $\text{DM} = 1000 \, \text{pc
  cm}^{-3}$. The observing frequency range is 400 -- 800 MHz, with
2048 frequency bins. The simulated data $I(t, f)$ is shown in the top panel
of \autoref{fig:sim}. The simulated data is truncated with a threshold 
$\tau = 3.0$. The truncated data $I_{m}$ is very sparse, as shown in the bottom panel of \autoref{fig:sim}, 
but the signal are mostly preserved.

\begin{figure}
  \centering
  \includegraphics[trim=5 10 40 34,clip,width=0.38\textwidth]{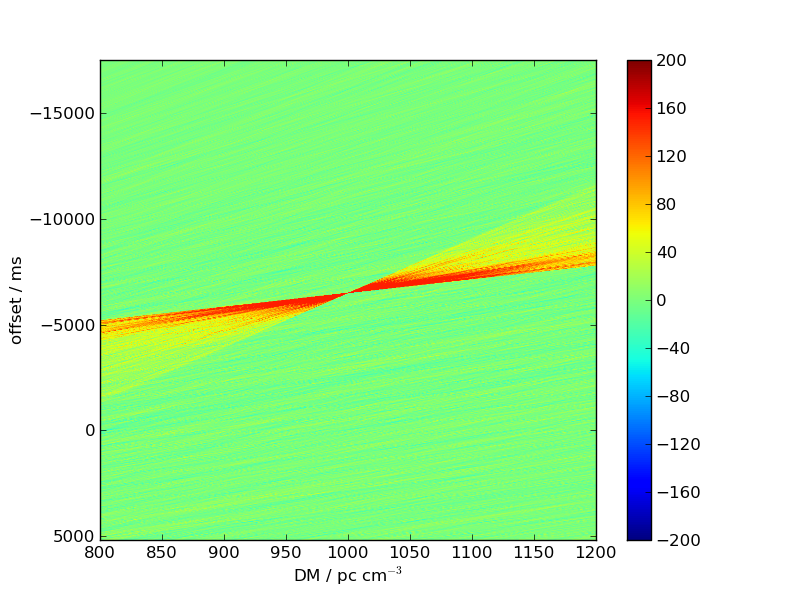}
  \includegraphics[trim=80 40 80 60,clip,width=0.38\textwidth]{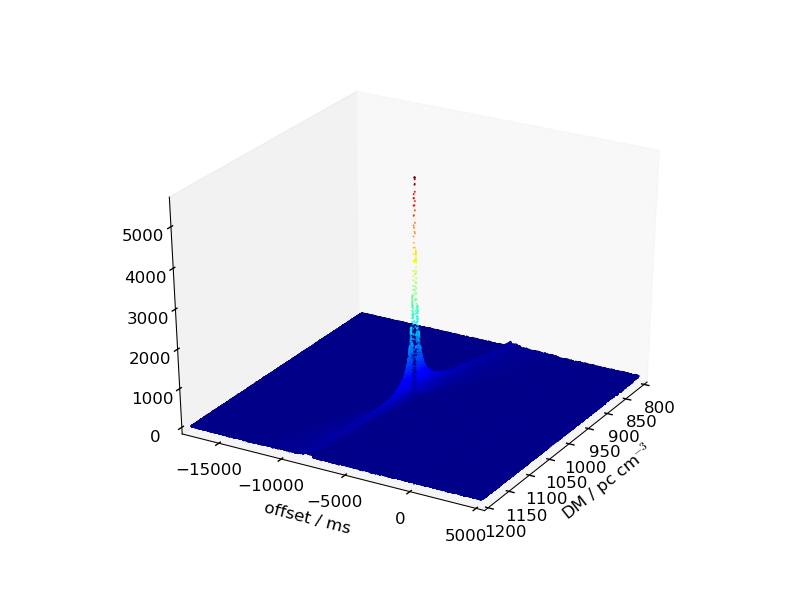}
  \caption{The Hough transform of the data shown in 2d (top) and 3d (bottom) plot. 
   }  
  \label{fig:peak}
\end{figure}

The Hough transform of the truncated data is show in the range 
of ${\rm DM} \in [800, 1200] \, \text{pc cm}^{-3}$
in \autoref{fig:peak}, from which we see the peak
is just at the right location $\text{DM} = 1000 \, \text{pc cm}^{-3}$. Here
we show both a 2D colour plot and a 3D plot for better illustration, as the strongest
point in the figure is too narrow that it is hard to see in the 2D plot.

\begin{figure}
 \centering
  \includegraphics[width=0.38\textwidth]{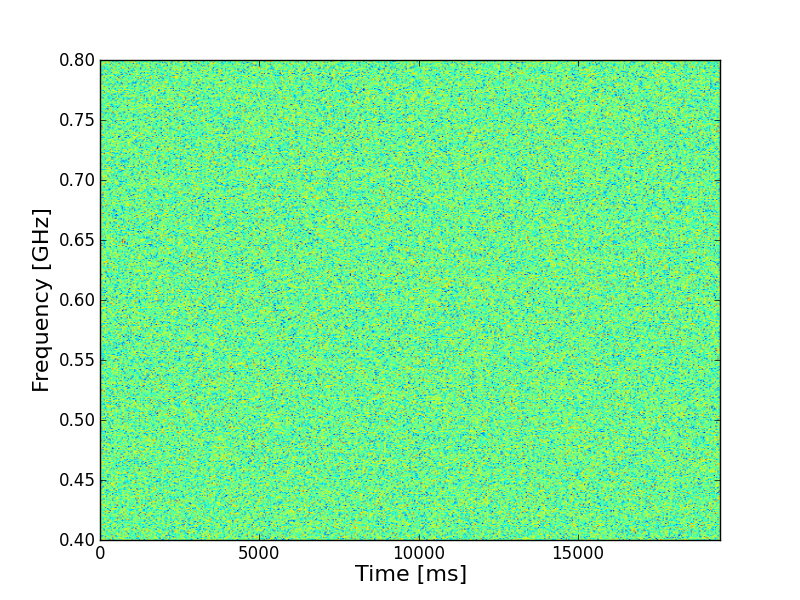}
  \includegraphics[trim=5 10 40 34,clip,width=0.38\textwidth]{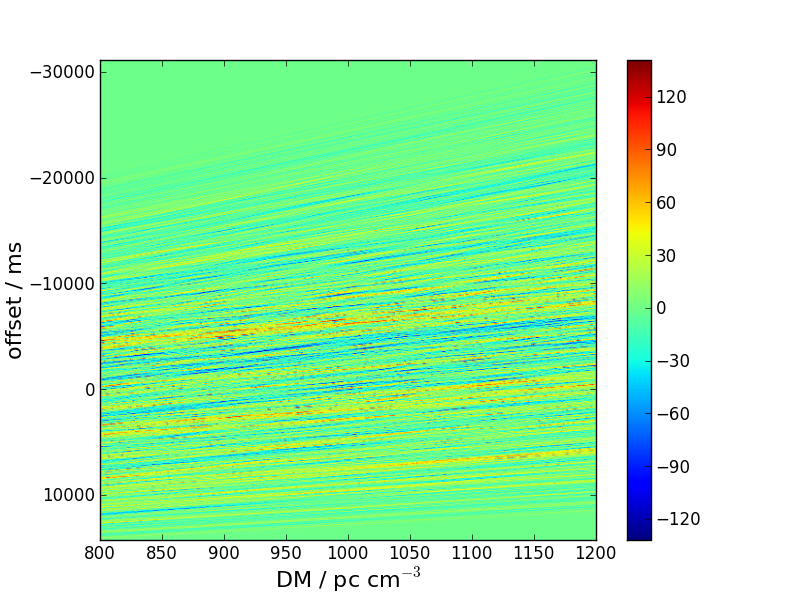}
 \caption{Top: A simulated noise data frame with $N_t=N_f=512$. Bottom:
 The Hough transform with no truncation.}
  \label{fig:simHough}
\end{figure}

To understand the pulse detection sensitivity in the Hough transform, we consider the distribution in the Hough transform 
 accumulator $A$. Here we first consider an ideal case, where the passband is flat, and the noise is uncorrelated white noise
 with Gaussian distribution. Here for definiteness, we consider a case where $N_t = N_f =512$, and $\sigma_n=1$, without 
 any pulse signal. The distribution of $A$ for this pure noise case can be simulated easily with a random number generator. 
 A typical data frame and its Hough transform with no truncation is shown in Fig.\ref{fig:simHough}. Each point in the time-frequency 
 data frame is mapped to a line in the Hough transform, so we see many line features in the bottom panel of Fig.\ref{fig:simHough}.

\begin{figure}
 \centering
  \includegraphics[width=0.38\textwidth]{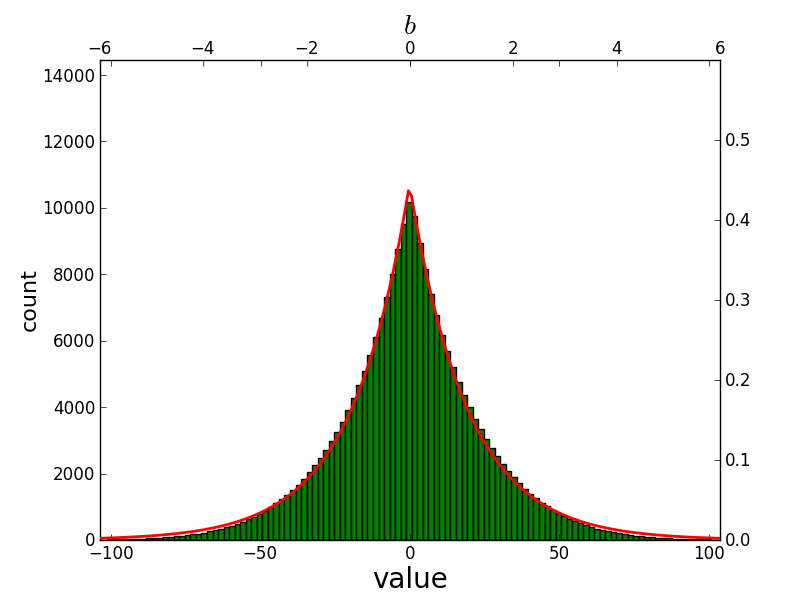}
  \includegraphics[width=0.38\textwidth]{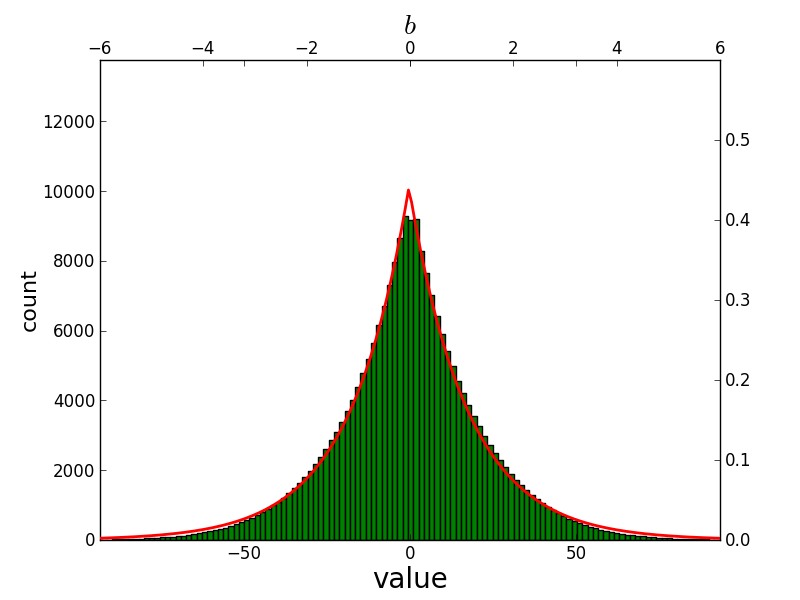}
    \includegraphics[width=0.38\textwidth]{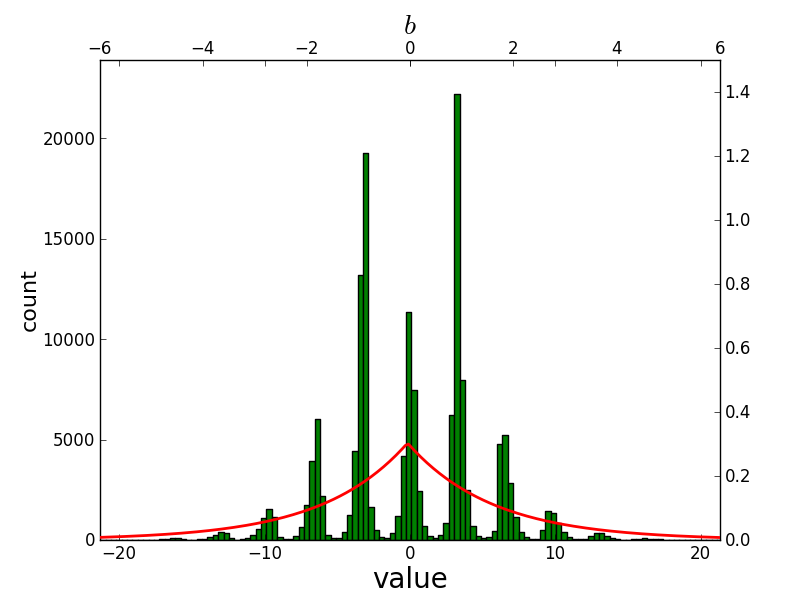}
 \caption{The Hough transform for a random noise image with threshold
  $\tau=0$ (top) and $\tau=1$ (middle) and $\tau=3$(bottom).}
  \label{fig:Adist}
\end{figure}

We then compute the distribution function of $A$, and show the results in  Fig.~\ref{fig:Adist}. 
 From top to bottom the panels are $\tau=0$ (no truncation), $\tau=1$ (low truncation threshold), and  $\tau=3$ (high truncation threshold) respectively.  For the no truncation case ($\tau=0$) or low threshold case
( $\tau=1$), the distribution can be well fitted by a Laplace distribution:
 \begin{equation}
 f_L(A |\mu, b) = \frac{1}{2b} \exp\left(-\frac{|A-\mu|}{b}\right).
 \end{equation}

The distribution in the $\tau=3$ case exhibits an interesting behavior: evenly spaced peaks appear in the plot. 
These peaks originates from the discrete nature of the data: with high truncation threshold, only the relatively 
 rare large fluctuation points remain in the time-frequency data frame. Each such point is transformed to a line in the 
Hough transform space, and contributes to a particular value of the elements of $A$ that the lines pass. According to Eq.~(\ref{eq:Aa}), this contribution is also proportional to the value $I(t, f)$, which is a real number, 
i.e. has a continuous distribution, so naively one might think that any discreteness would be washed out and 
the distribution should be smooth. However, because the Gaussian distribution function of the noise is 
 quite steep, when a particular high threshold is chosen, most survival points would be just above the threshold and 
 have similar values. For example, for the truncation threshold $\tau = 3$, most of the survival points would have a value 
 $I_{p}$ just above $3$, they all map to lines with this intensity in $A$, and produce the first non-zero peak in the distribution of 
 $A$. The crossing of these lines contributes a multiples of $I_{p}$ to the accumulator, so in the end we see these 
 evenly spaced peaks in the  $A$ distribution.

%

Based on the $A$ distribution function, one can choose a detection threshold for $A$, such that a false detection rate is 
below a preset value. For example, if we take the threshold to be $|A-\mu|>4b$, where $b$ is the parameter in the best-fit Laplace 
distribution, the false detection rate falls below 1.8\%, or 0.9\% if we only consider the positive end. 

However, in the real application the situation would be much more complicated, as the passband may not
be flat, the noise could be correlated, and there are also radio sources and RFIs. One will have to study the specific 
instrument and derive the noise generated $A$ distribution from the data.

\begin{figure}
 \centering
  \includegraphics[trim=10 10 40 34,clip,width=0.38\textwidth]{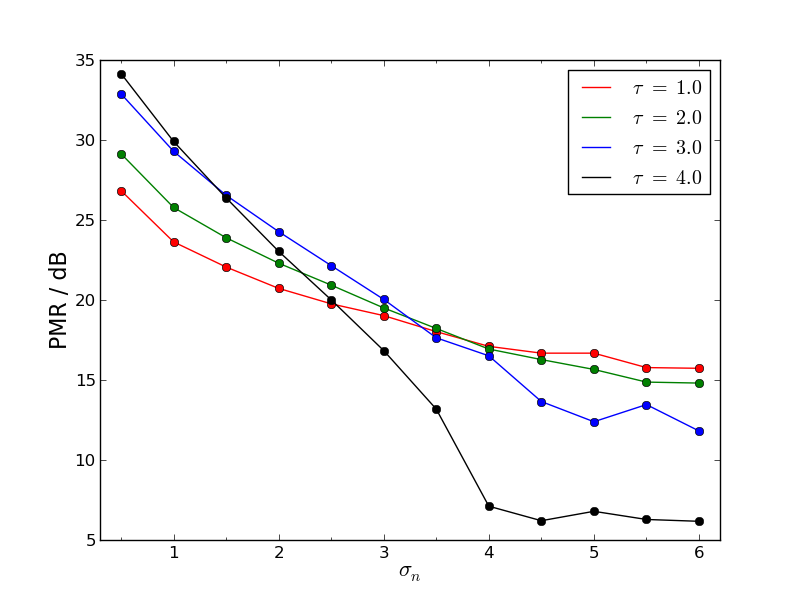}
    \includegraphics[trim=10 10 40 34,clip,width=0.38\textwidth]{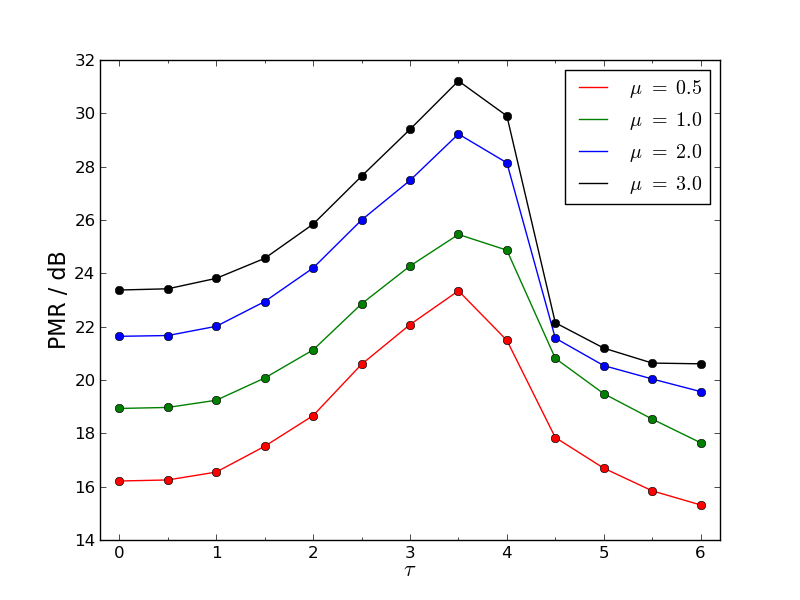}
 \caption{Top: the Peak-to-Median Ratio (PMR) shown in dB for different noise level $\sigma_n$, Bottom: the 
 PMR (shown in dB) for different threshold $\tau$.}
  \label{fig:pmr}
\end{figure}


The Peak-to-Median Ratio  $\text{PMR} = \max(A) /\text{median}(A_{> 0})$ can be used as 
a measure of the relative strength of the highest peak to the noise background level, 
where $A_{> 0}$ is the set of points for which the accumulator $A$
have positive values.
The PMR as a function of the noise level $\sigma_n$ (with $\tau=1,2,3,4$) are plotted in the top panel of
 \autoref{fig:pmr} (shown in dB scale, i.e. $10\log_{10} {\rm PMR}$).   We see that generally the PMR decreases monotonically as the noise level 
 $\sigma_{n}$ increases, but up to $\sigma_{n} = 6.0$ the overall PMR is still sufficiently high
 (PMR=14.3, or 11.5 dB) for detection. We also plot curves for different threshold value $\tau$. 
 A higher threshold $\tau$ generally yields better SNR at low noise level $\sigma_n$,
 but at high noise level $\sigma_n$ this is reversed. 


In the Bottom panel of \autoref{fig:pmr} we plot the PMR (shown in dB scale) as a function of the 
selected threshold $\tau$, for fixed noise level but 
different mean values. Initially the PMR value increases as $\tau$ increases, 
but then it decreases after reaching a peak.  The truncation 
process does not only reduces computation complexity, but also throws away the relatively noisy data, 
while preserving the data where the signal is stronger, so the PMR is
enhanced for some appropriate $\tau$, and a highest PMR is achieved at $\tau \sim 3.5$.
The effect of truncation threshold may also depend on the mean background level. Several different
mean value $\mu$ are plotted. However, as the curves show, although the peak value of PMR depends 
strongly on the $\mu$ value, for low $\mu$ the peak PMR is small (e.g. when $\mu = 0.5$, we
can get a peak PMR of 117.9(20.7 dB), while for $\mu=3$ the peak PMR 
is  up to $\sim 1350$(31.3 dB). However, 
the value that achives highest PMR does not change much.  Even in the case
when $\mu = 0$, as long as $\sigma_{s}$ is much higher than
$\sigma_{n}$, the pulse can be detected by the Hough transform
method, for example, when $\mu = 0$, $\sigma_{s} = 3.0$, while
$\sigma_{n} = 1.0$, we obtain a PMR of 19.6 (12.9 dB) when using a truncation
threshold $\tau = 3.0$. 

These results show that the Hough transform
pulse detection method works quite well for
  relatively strong S/N pulse signals. However, the selection of the truncation threshold is 
  a delicate problem, a lower threshold is needed to detect the low signal-to-noise ratio 
  pulses, but the computation time increases drastically at lower threshold. There is
  a trade-off between the computation savings and the mis-detection
  rate, one needs to select the threshold according to practical needs.

\begin{table*}
  \centering
  \begin{tabular}{llllll}
    \hline\hline
    Event   & Telescope   & DM [pc cm$^{-3}$]  & S$_{\rm peak,obs}$ [Jy]  & F$_{\rm obs}$ [Jy ms]  & Ref \\
    \hline
    FRB~010125  & Parkes  & 790(3)  & 0.30  &  2.82 & \cite{Spolaor2014} \\
    FRB~010621  & Parkes  & 745(10)  & 0.41  &  2.87 & \cite{Keane2011} \\
    FRB~010724  & Parkes  & 375  & {\small$>$}$30.00^{+10.00}_{-10.00}$  & {\small$>$}$150.00$ & \cite{Lorimer2007} \\
    FRB~110220  & Parkes  & 944.38(5)  & $1.30^{+0.00}_{-0.00}$ & $7.28^{+0.13}_{0.13}$  & \cite{Thornton2013} \\
    FRB~110523  & GBT  &  623.30(6) & 0.60 & 1.04  & \cite{Masui2015} \\
    FRB~110626  & Parkes  & 723.0(3) & 0.40 & 0.56  & \cite{Thornton2013} \\
    FRB~110703  & Parkes  & 1103.6(7) & 0.50 & 2.15  & \cite{Thornton2013} \\
    FRB~120127  & Parkes  & 553.3(3) & 0.50 & 0.55  & \cite{Thornton2013} \\
    FRB~140514  & Parkes  & 562.7(6) & $0.47^{+0.11}_{-0.08}$ & $1.32^{+2.34}_{-0.50}$  & \cite{Petroff2015} \\
    \hline\hline 
  \end{tabular}
  \caption{FRBs used in this paper and some of their
    parameters.}
  \label{tab:frbs}
\end{table*}

\begin{figure*}
  \centering
  \includegraphics[trim=10 10 40 34,clip,width=0.29\textwidth]{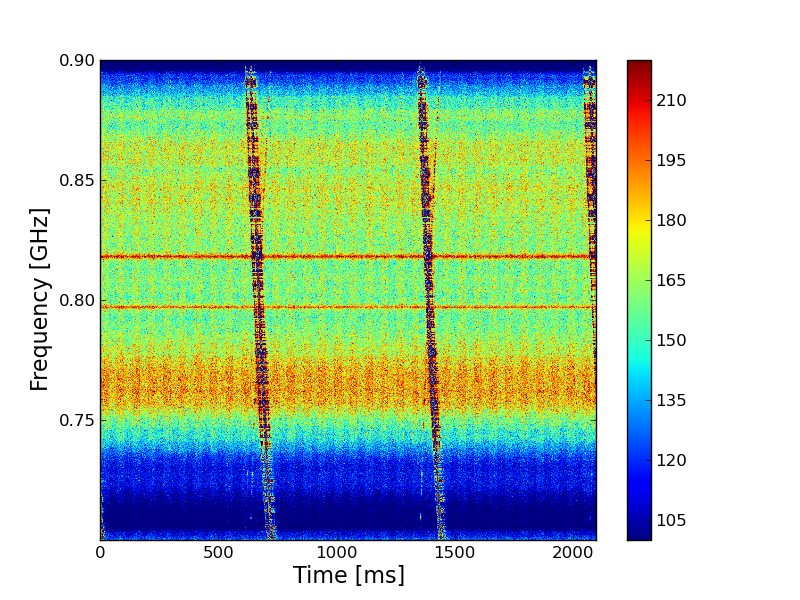}
  \includegraphics[trim=10 10 40 34,clip,width=0.29\textwidth]{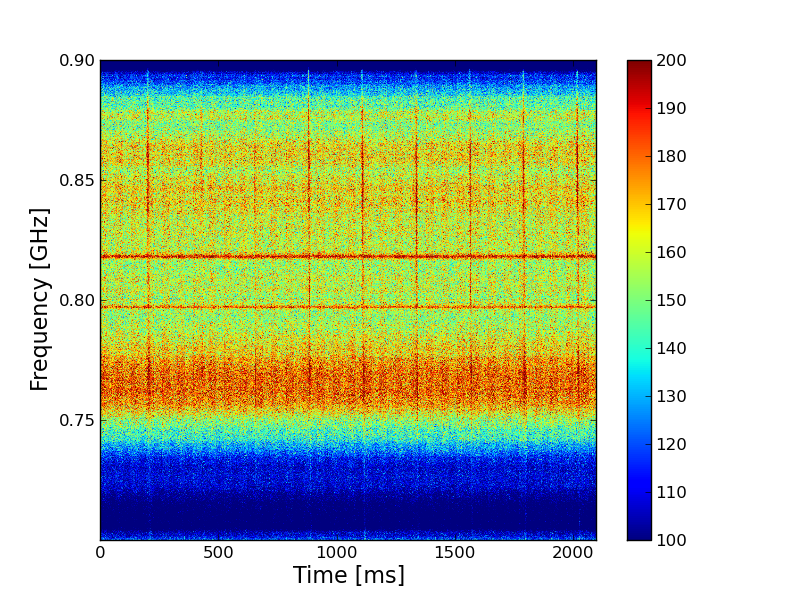}
  \includegraphics[trim=10 10 40 34,clip,width=0.29\textwidth]{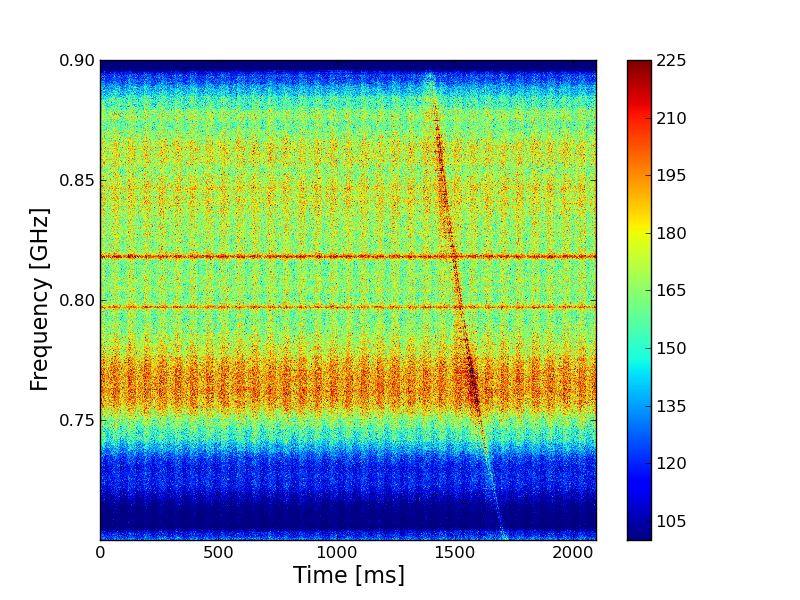} \\
  \includegraphics[trim=10 5 40 34,clip,width=0.29\textwidth]{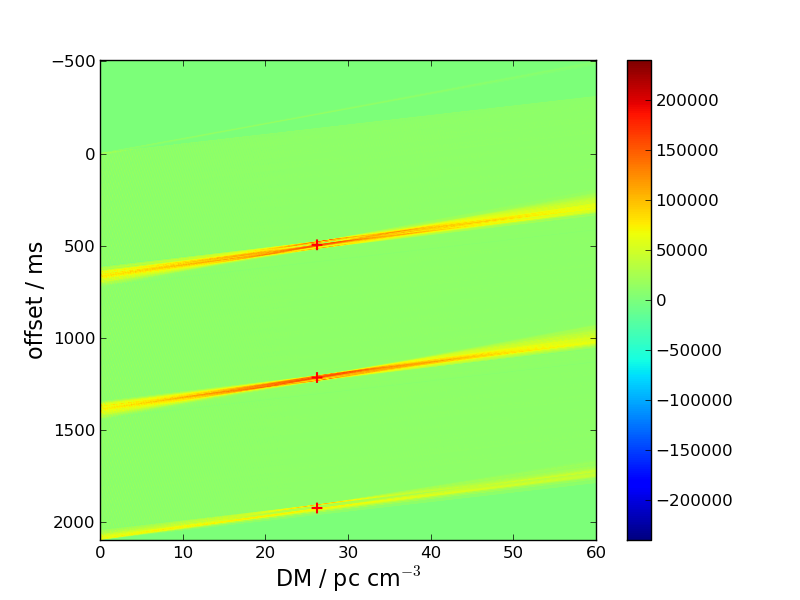}
  \includegraphics[trim=10 5 40 34,clip,width=0.29\textwidth]{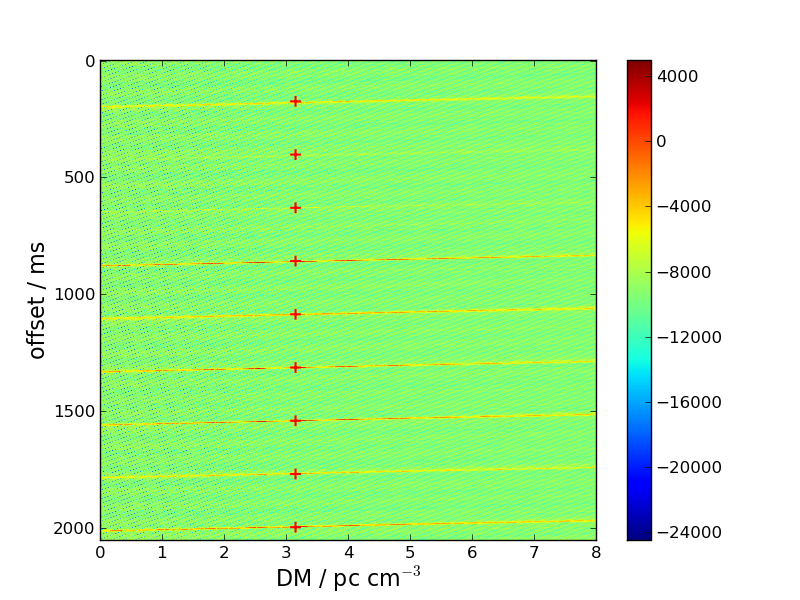}
  \includegraphics[trim=10 5 40 34,clip,width=0.29\textwidth]{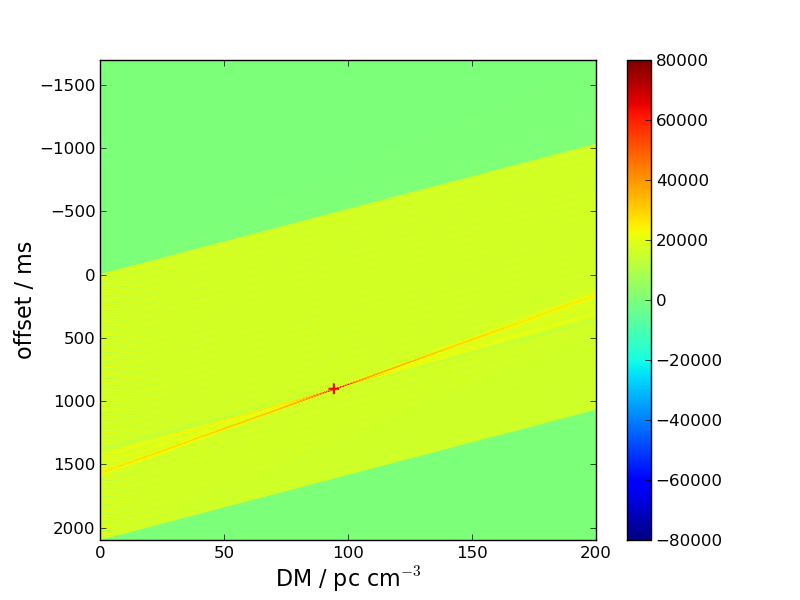}
  \caption{The observing data (top) and the corresponding Hough transform (bottom) of
    pulsars B0329+54, B1929+10, and B2319+60 respectively. The detected peaks are marked by
  a red $+$ in the transformed image.}
  \label{fig:psr}
\end{figure*}

\subsection{Real Pulsar and FRB Data}\label{S:rd}

We first apply our method to real observation data of three pulsars, i.e.
B0329+54, 
B1929+10, 
and B2319+60 
taken by the Green Bank Telescope (GBT)\footnote{\url{https://dss.gb.nrao.edu/project/GBT14B-339/public}}.
We chose a truncation threshold $\tau = 3.0$, and used a $4000 \times 2000$ 
accumulator $A(t_{0}, d)$ with a DM range $[0, 100] \, \text{pc cm}^{-3}$. The Hough transform for 
an example of the data lasting about 2 seconds  containing several pulses are shown in the top panels of \autoref{fig:psr} for the 
three pulsars. The corresponding Hough transform are plotted in the bottom panels. 
and we have also marked the detected peaks by a red $+$ in the transformed images.

 In the data shown in the plot,  there is one single pulse track for B2319+60 (right column) with 
 DM $ =94.591 \text{pc cm}^{-3}$, while for B0329+54 (left column) there are three tracks with 
 DM $ =26.7641 \text{pc cm}^{-3}$, and many tracks for B1929+10 (middle row) with 
 DM $ =3.18321 \text{pc cm}^{-3}$.  Each pulse track in the observed
data has been transformed to a bundle of lines crossing at the same
point, which can be detected easily with the program (though when plotted in \autoref{fig:psr}
they are visually not so obvious due to the small size of the peak point, to aid the eye we 
marked these by a cross in the figure.)
If there are more than one cross points corresponding to more
than one pulse track in the observed data, they all have the same DM
value, which are all very close to the values measured with other programs (e.g.
those given by {\tt psrcat} in \cite{Manchester2005}).
Note there are two strong narrow frequency band RFIs
in the middle of each data, but they do not show in the
corresponding Hough transformed images, as the Hough transform automatically rejects them.

\begin{figure*}
  \centering
  \includegraphics[trim=10 10 40 34,clip,width=0.29\textwidth]{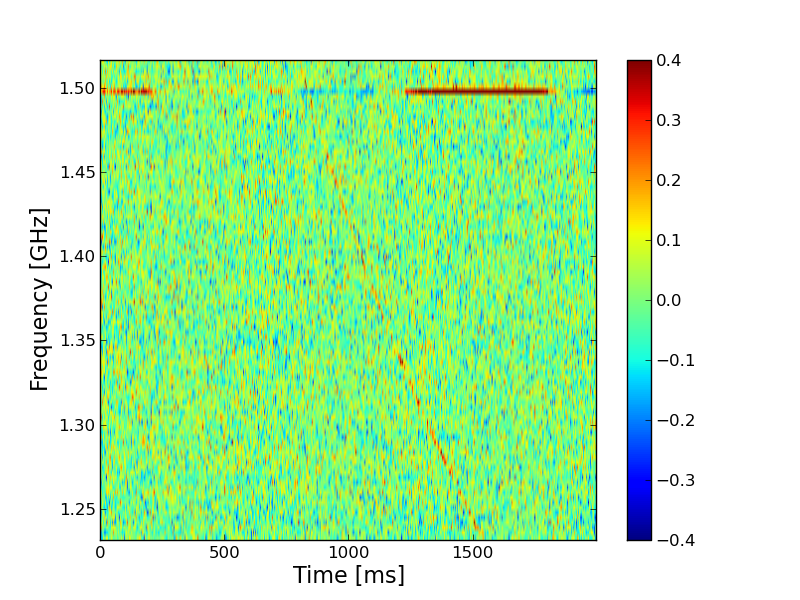}
  \includegraphics[trim=10 10 40 34,clip,width=0.29\textwidth]{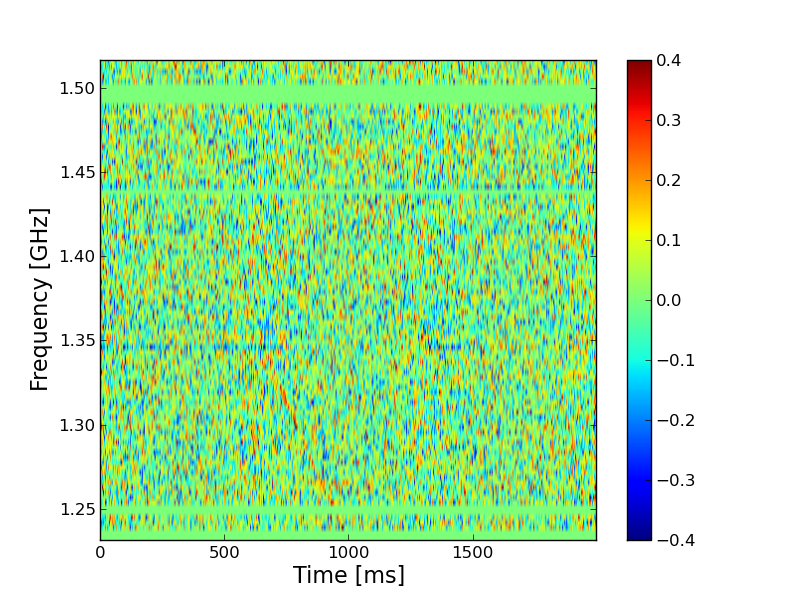}
  \includegraphics[trim=10 10 40 34,clip,width=0.29\textwidth]{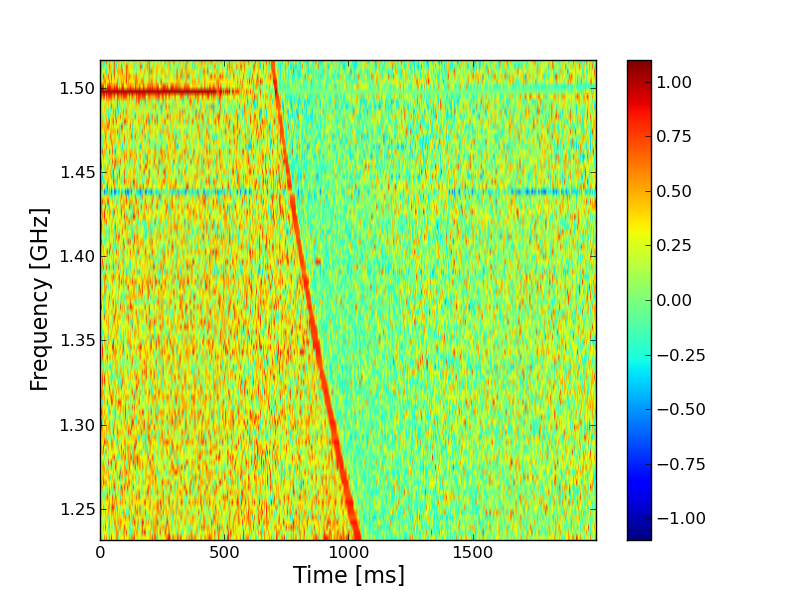} \\
  \includegraphics[trim=10 10 40 34,clip,width=0.29\textwidth]{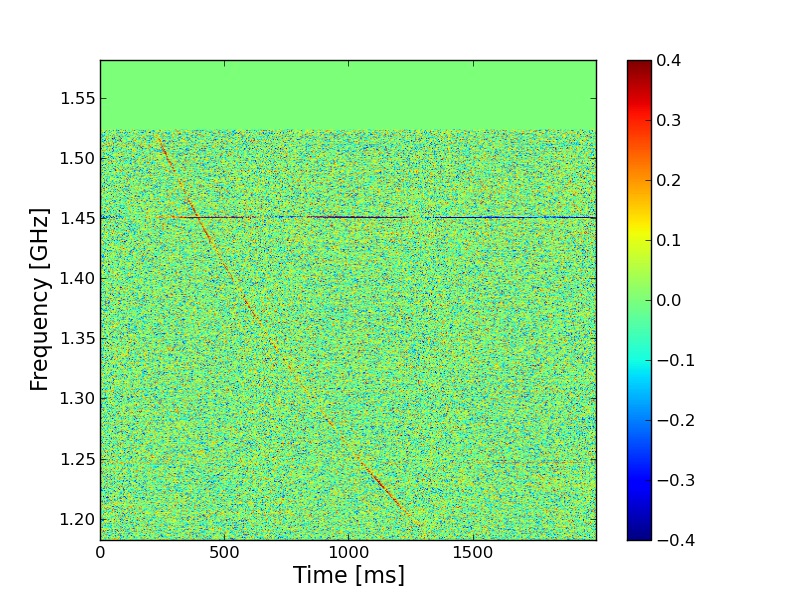}
  \includegraphics[trim=10 10 40 34,clip,width=0.29\textwidth]{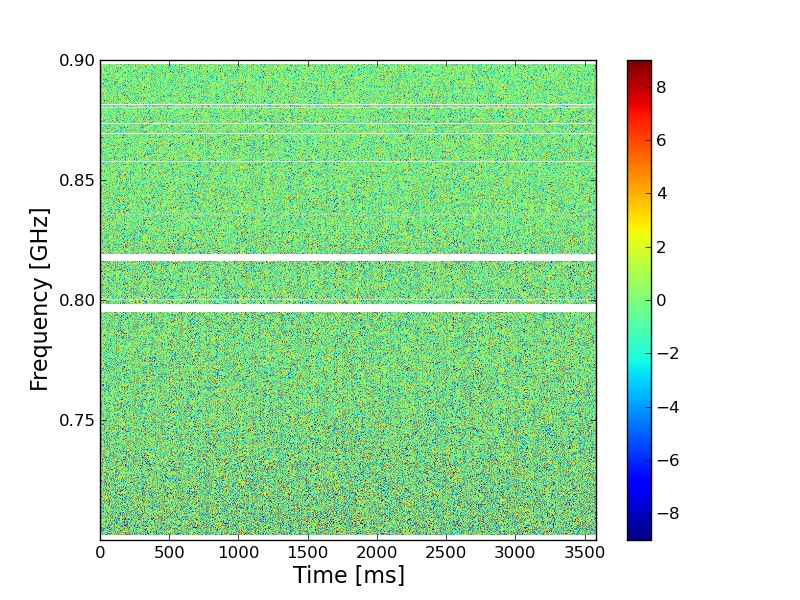}
  \includegraphics[trim=10 10 40 34,clip,width=0.29\textwidth]{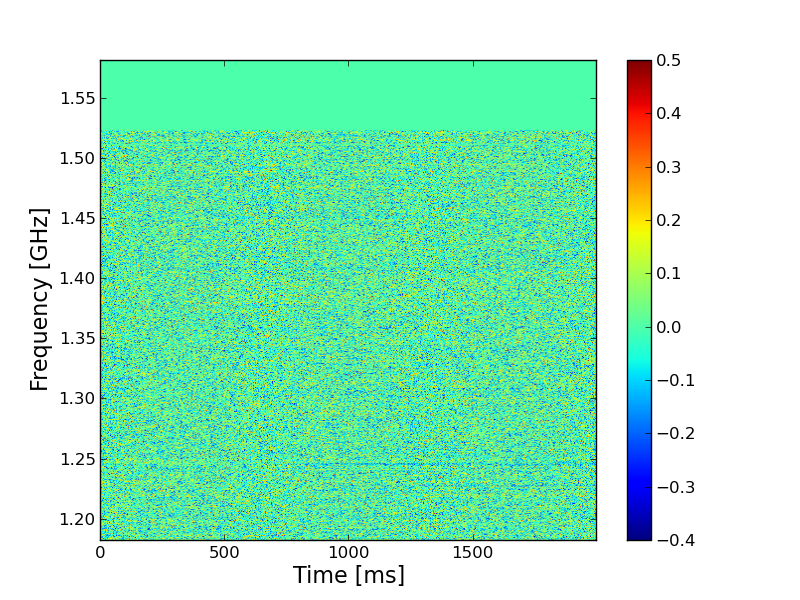} \\
  \includegraphics[trim=10 10 40 34,clip,width=0.29\textwidth]{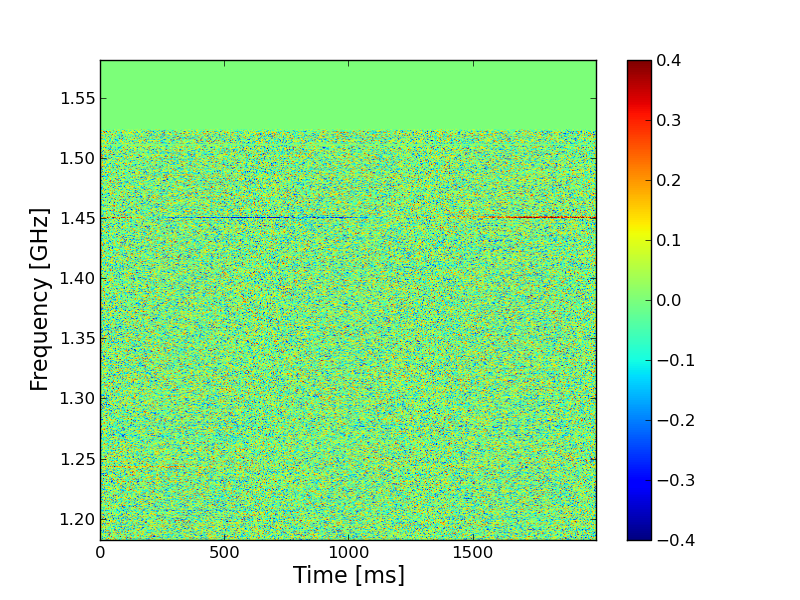}
  \includegraphics[trim=10 10 40 34,clip,width=0.29\textwidth]{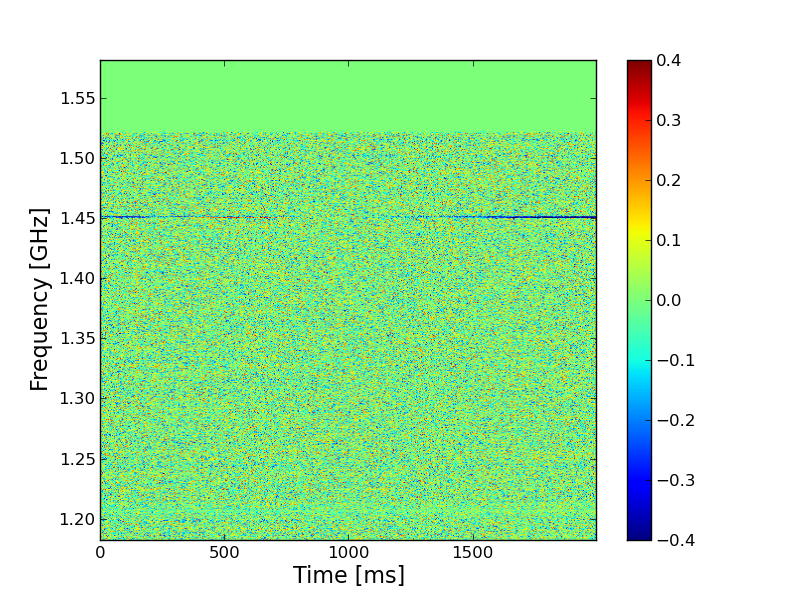}
  \includegraphics[trim=10 10 40 34,clip,width=0.29\textwidth]{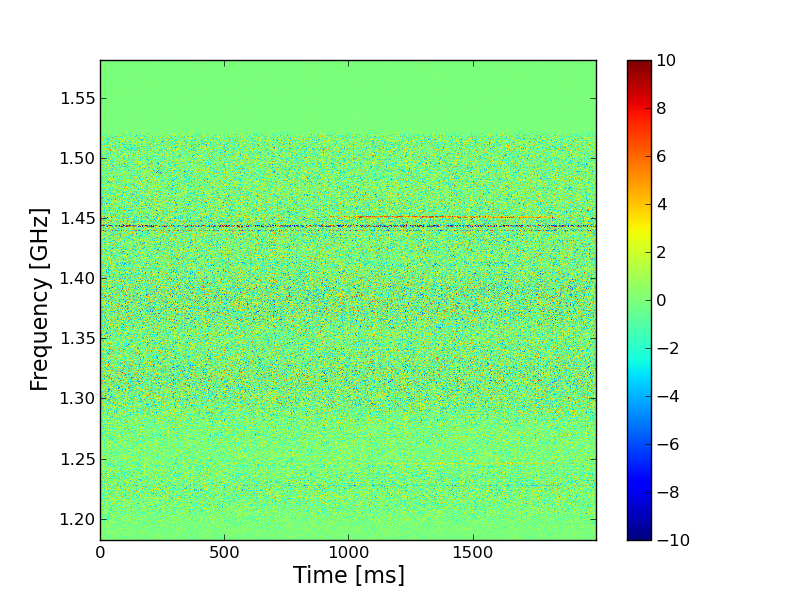} \\
\caption{The observation data for FRB~010125, FRB~010621, FRB~010724(top row, left to right), 
 FRB~110220, FRB~110523, FRB~110626 (middle row, left to right), and FRB~110703, FRB~120127, FRB~140514(bottom row, 
 left to right).}
   \label{fig:frbI}
\end{figure*}

\begin{figure*}
  \centering
  \includegraphics[trim=10 5 40 34,clip,width=0.29\textwidth]{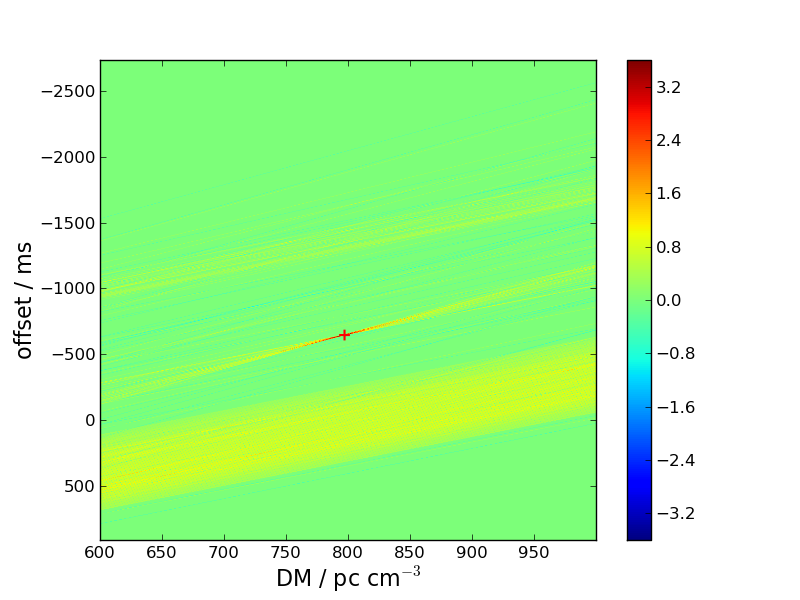}
  \includegraphics[trim=10 5 40 34,clip,width=0.29\textwidth]{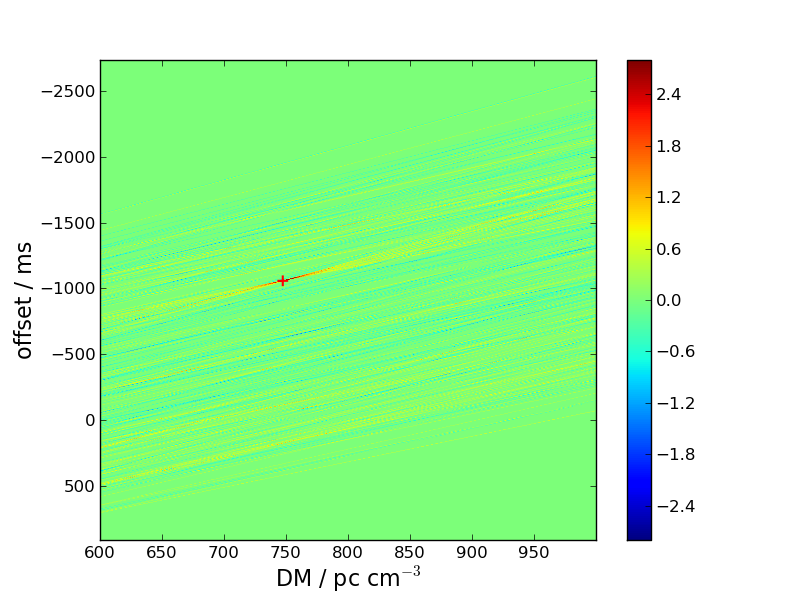}
  \includegraphics[trim=10 5 40 34,clip,width=0.29\textwidth]{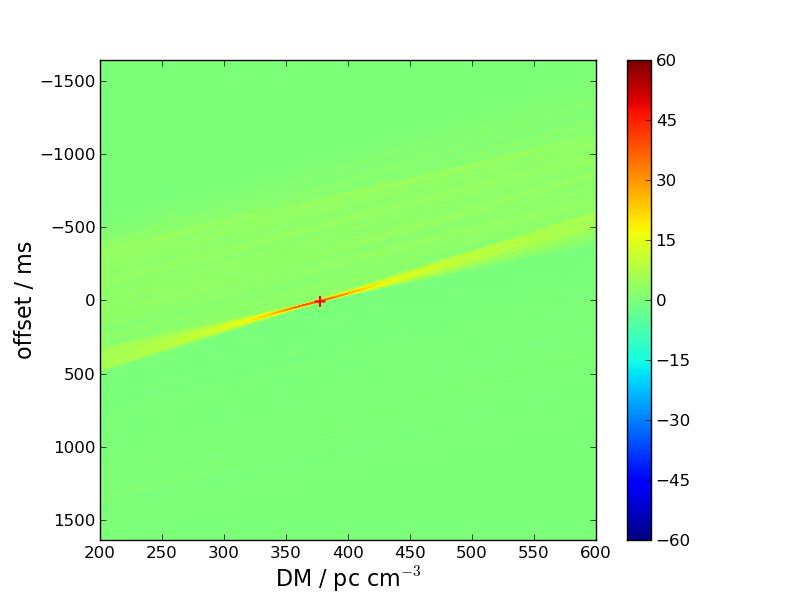} \\
  \includegraphics[trim=10 5 40 34,clip,width=0.29\textwidth]{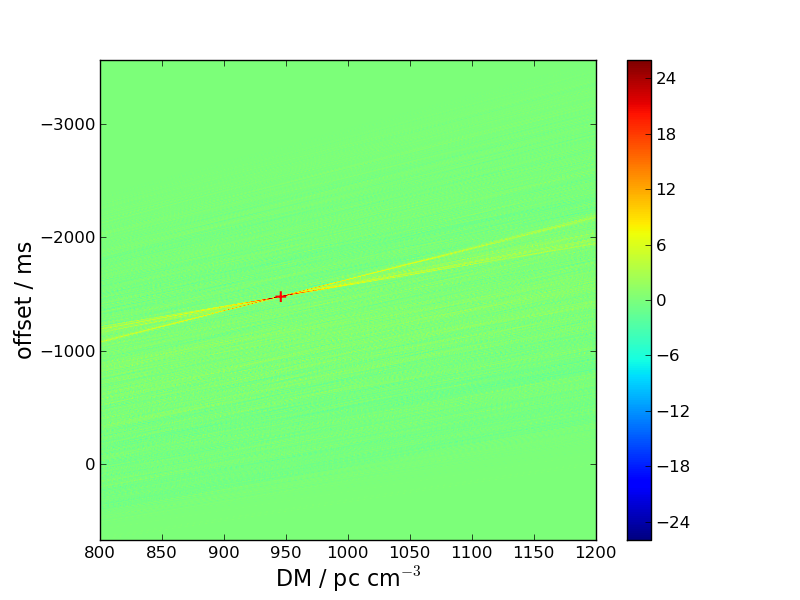}
  \includegraphics[trim=10 5 40 34,clip,width=0.29\textwidth]{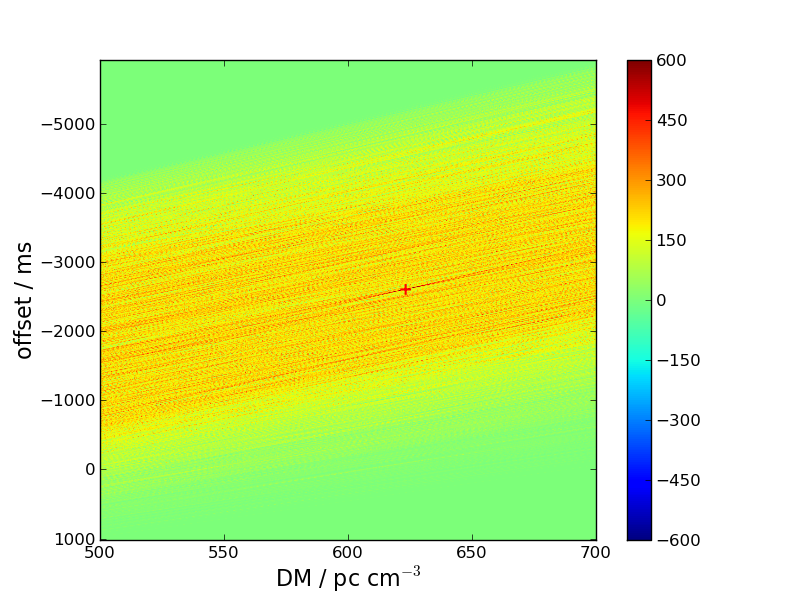}
  \includegraphics[trim=10 5 40 34,clip,width=0.29\textwidth]{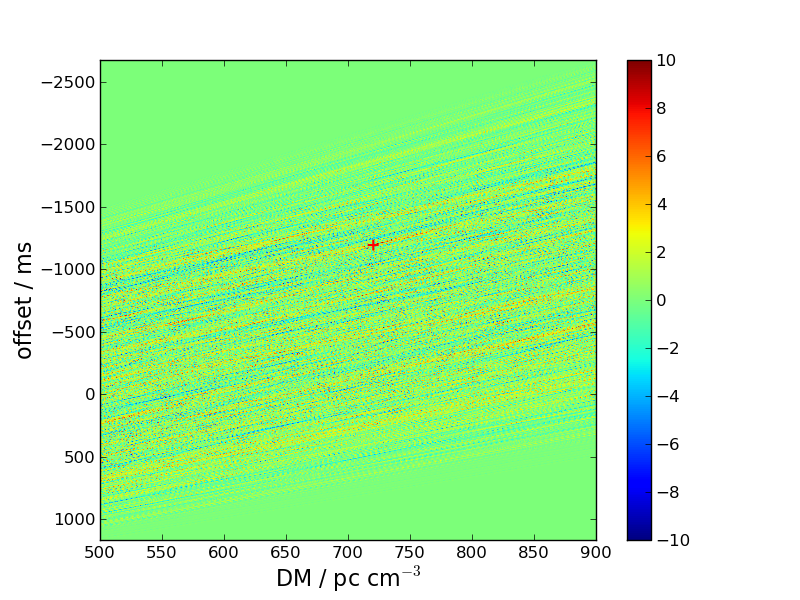} \\
  \includegraphics[trim=10 5 40 34,clip,width=0.29\textwidth]{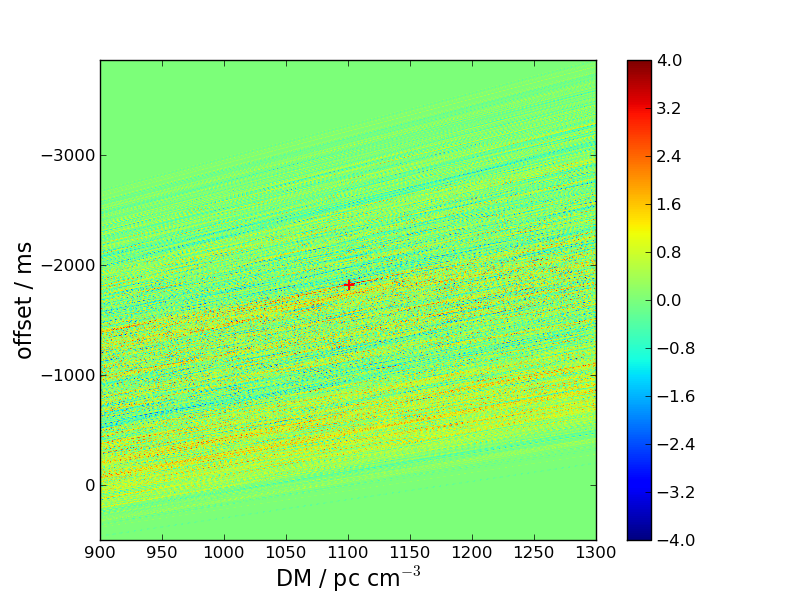}
  \includegraphics[trim=10 5 40 34,clip,width=0.29\textwidth]{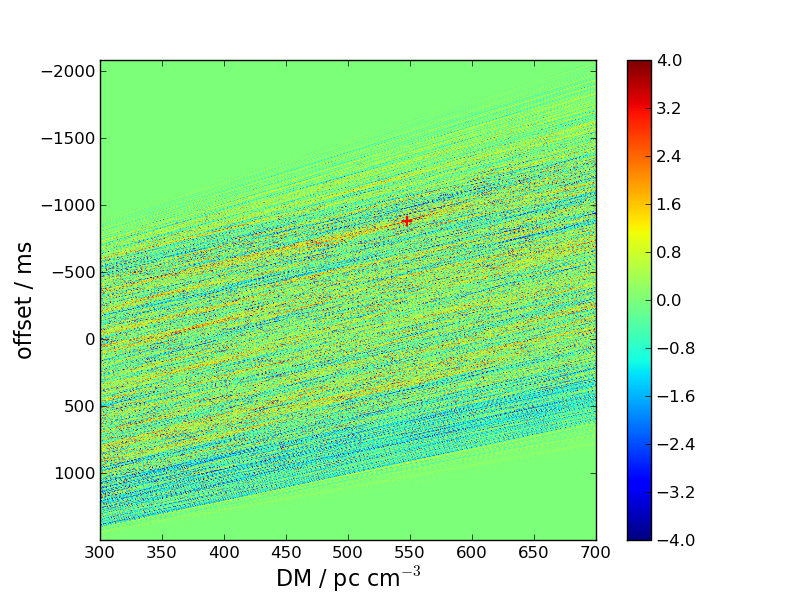}
  \includegraphics[trim=10 5 40 34,clip,width=0.29\textwidth]{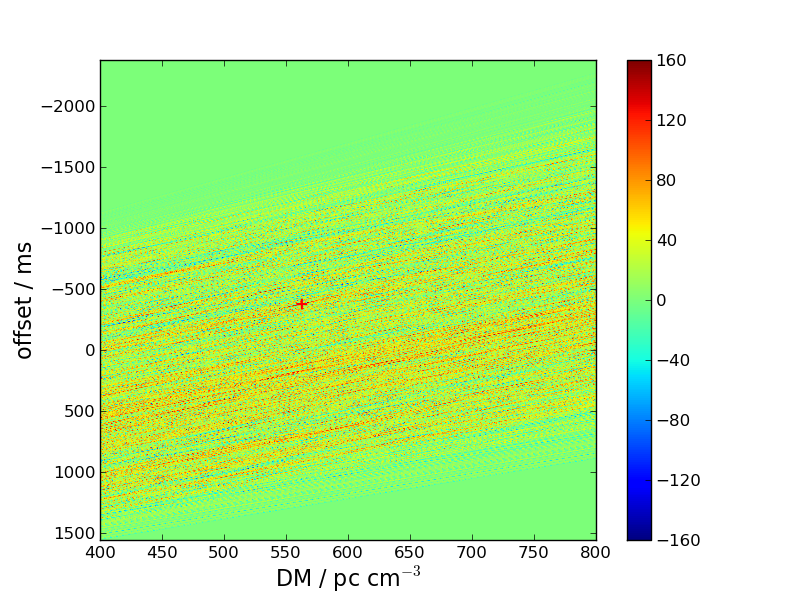}
  \caption{The Hough transform matrix $A$ for the 9 FRBs shown in the figure above, in the 
  same order. The detected peaks are marked by a red $+$ in the transformed image.
  }
  \label{fig:frbA}
\end{figure*}

We also apply our method to several FRB event data, including eight 
FRBs observed by the Parkes telescope taken from the FRB Catalogue\footnote{\url{http://www.astronomy.swin.edu.au/pulsar/frbcat/}}
 compiled by \cite{Petroff2016}, and 
one FRB event data (FRB~110523) \footnote{\url{http://www.cita.utoronto.ca/~kiyo/release/FRB110523/}} observed by the 
 GBT \citep{Masui2015}.  These FRBs are listed in \autoref{tab:frbs}. 
 Observing data of these FRBs are shown in Fig.\ref{fig:frbI}, 
and their corresponding Hough transformed result are shown in Fig.\ref{fig:frbA},
with the detected peaks marked by red $+$ signs in the transformed image. 


We tried a few different truncation thresholds, then apply the Hough transform to search for them. 
We found that in most cases  a threshold $\tau = 3.0$ is sufficient, but for a few weaker ones 
lower thresholds are required,  specifically,
$\tau = 2.5$ for FRB~120127, $\tau = 2.0$ for FRB~010724, and $\tau
= 1.0$ for FRB~110626. 
We see from \autoref{tab:frbs} that FRB~120127 and FRB~110626 have very
low peak intensity S$_{\rm peak,obs}$ and integrated intensity F$_{\rm obs}$ relative
to other ones, lower thresholds are needed for their detection. The lower threshold would
require larger amount of computation,  while in higher threshold they might be missed. 
The case of FRB~010724 is 
somewhat different, its pulse signal is fairly strong, but it could also only be detected with a lower
threshold, say $\tau= 2.0$.  This may be related to its non-uniform background
noise, as can be seen clearly in \autoref{fig:frbI}, the background
noise has big difference between the left and right part of the
pulse track, this affects the background mean subtraction, and
further makes the pulse detection harder. Nevertheless, 
all FRBs can be successfully detected by using 
the Hough transform method, and the DM values obtained from the 
detected peaks are very close to the public values
as listed in \autoref{tab:frbs}.

Note that in our processing, we have not make any special treatment for
the RFIs or any other outliers before the Hough transform. 
For all data except for the FRB~110523, we have used the raw data, the only 
processing besides the truncation and Hough transform described above is 
rebinning in time direction to reduce the amount of data. For FRB~110523
the pre-processed data is available to us, which has been calibrated and RFI flagged. 
We can see from the data images, many of the data has RFIs or outliers in
them, usually single frequency or narrow band RFIs, some are much stronger than the pulse signals.
As we discussed in Section.~\ref{S:disc}, they should not have much impact on the detection based on the 
Hough transform,  and this is confirmed by the results. 

In the present treatment, the search for the peak in $A$ is
conducted at the single pixel level, i.e. we search the pixel which
has the maximum value. This may not achieve the highest sensitivity, for the true peak
may be located somewhere between the accumulator pixels, so that each of its neighbouring pixels 
has a relatively high value, but not as high as it would have been if the peak is right centered on that pixel.

A related issue is, in the above we have assumed that the time width of the pulse signal is within one sample, however, 
the burst may last more than one sample width, i.e. the signal track is not a curve but a band of finite width. 
In such cases, the Hough transform will map the band into several neighbouring pixels in the accumulator space, i.e. $A(d,t_0)$
of the same $d$ but successive $t_0$. For optimal detection sensitivity, the width of the 
signal should be nearly equal to the sample time width. A practical implementation of pulse search program needs to 
deal with this issue. In some search programs, this can be achieved by rebinning the data along the time direction,
 up to a physically plausible maximum width, and make trial search with the larger sample time widths. In the Hough 
 transform, this process can be achieved simply by integrating the neighbouring pixels with kernels of different widths. 
 Here the Hough transform may also have some advantages over other methods, for instead of blindly rebinning 
 along time direction for all data points, here we may pick a small number of peaks in the accumulator space, and 
 try integrating with kernels of different widths, in practice make weighted average with different weights.
 The amount of  computation in this part is again much reduced compared with rebinning all points. This is also
 more flexible than rebinning, and can even  be made adaptively. We shall leave the detailed investigation 
 of these issues to future studies.

\section{Discussions}\label{S:conc}
We have presented a simple and fast radio bursts detection and
incoherence dedispersion method based on Hough transform. The
$f^{-2}$ burst curve in the observed time stream data is mapped to
be a bundle of straight line in the transformed space which
crossed the same point determined by the dispersion measure of the
burst. By detecting the peak, we can detect the bursts and measure their dispersion 
measures. The advantage of the method is that by setting an appropriate truncation threshold, it has
a low computational complexity of $O(\text{max}(N_{t}, N_{f}) N_{d})$, which is much lower than the 
existing algorithms. Wise choice of noise
truncation threshold may also improve the detection sensitivity. The method automatically rejects most 
commonly encountered RFIs, making it  good for online (real time) bursts detection. We
have shown its effectiveness by simulation and application to the
real pulsar and FRB observing data. 

\begin{figure}
 \centering
 \includegraphics[trim=10 10 40 34,clip,width=0.4\textwidth]{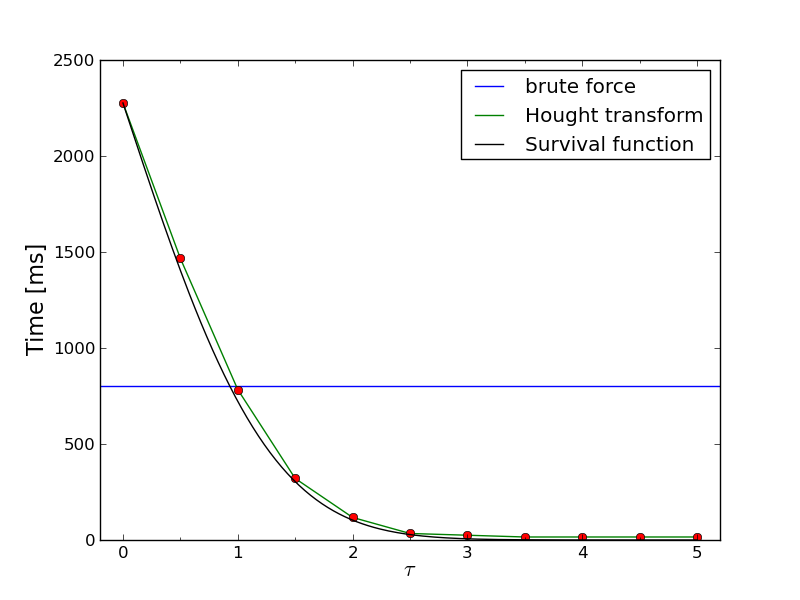}
 \includegraphics[trim=10 10 40 34,clip,width=0.4\textwidth]{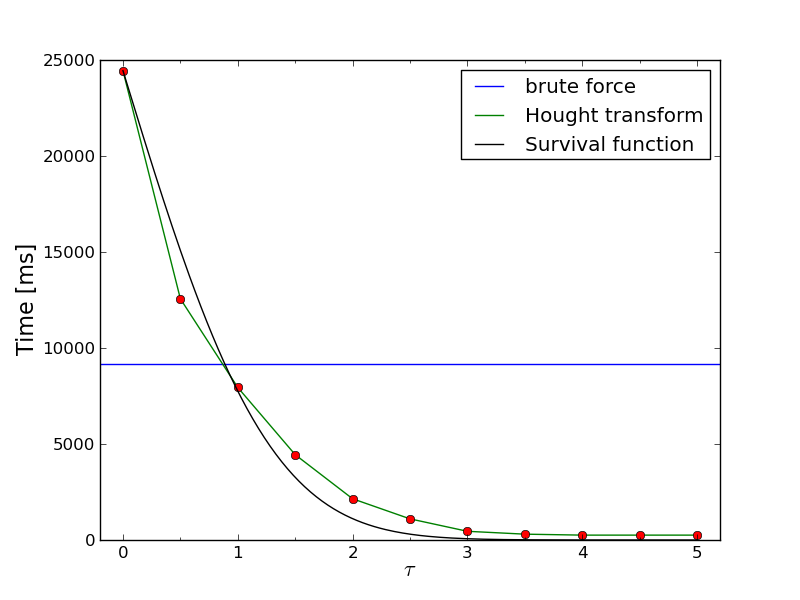}
 \caption{Comparison of the computation time for the Hough transform
 method with different truncation threshold $\tau$ and the brute
 force method. Top panel for FRB~010125, Bottom panel for FRB~110220.  
 The black curve gives the scaled survival function $2
  \Phi(-\tau) = 2 [1 - \Phi(\tau)]$ of the
  normal distribution, which overlaps the computation time curve at
  the value of $\tau = 0$.}
  \label{fig:tm}
\end{figure}

The truncation of data which saves much computation is particularly important for real-time detection. 
The computation time for several different truncation thresholds are shown 
in \autoref{fig:tm} for FRB~010125 with data dimension of $96 \times 500$ and 
for FRB~110220 with data dimension of $1024 \times 500$). The computation time are measured on 
an Intel Xeon E5-2670 2.60~GHz CPU using program written in the C programming language. For comparison, the
brute force dedispersion is also implemented in the same computing environment, the algorithms 
run with a single thread for the same range and resolution of dispersion measure, and the
time reported is the average time for 10 runs. Note that here we
  have included the background subtraction time in the reported
  computation time of the Hough transform method, where for the
computation of the median (and also the MAD), we have simply implemented it by first
sorting the array, which is not a very fast method for median
computation, more effective methods exists, for example, the median
of medians algorithm \citep{Blum1973}, which finds an approximate median in linear time. 
If some of the faster median computing methods are
used, the performance of the Hough transform method can be further
improved. But even for this simple implementation, the
computation time of the Hough transform method with a truncation
threshold $\tau \ge 1.0$ is less than that of the brute force
method, and the computation times decays exponentially with the
increasing truncate threshold $\tau$. The exponential fall off
is what we expected for the survival number of pixels by thresholding a Gaussian background. 
We also plot the survival function $2  \Phi(-\tau) = 2 [1 - \Phi(\tau)]$ of the
  normal distribution in \autoref{fig:tm} for comparison.
  
The Hough transform method is readily applicable to online (nearly
real time) processing. It can also be easily parallelized to
speed up the computation, by either partitioning the
points in the truncated image to $N$ parts and do the Hough transform
for these points in each part independently, with the total accumulator given by the sum, 
or by partitioning the computation of different dispersion
measures. The Hough transform algorithm itself can also be
  parallelized \citep{Tzvi1989,Lu2013}, or be accelerated with graphic processing units (GPUs) \citep{Tomagou2013,Patil2016}.
 GPUs  are now widely used in computing-intensive environment. With very large number 
of computing cores,  a GPU can achieve orders of magnitude higher computing speed than a CPU for an appropriate problem, 
by dividing the work load among thousands of threads which run in parallel.  Indeed, given the much higher 
raw computing speed,  even the brute force de-dispersion could be handled with the GPUs in short time. 
Nevertheless, improving the computing efficiency is still important, for the data rate is also  rapidly increasing. For example, 
thousands of beams can be formed by a large interferometer array such as the SKA, each with very large bandwidth, and 
searching the FRB for these many beams remain a computationally challenging task.

We now consider the Hough transform with GPU. 
 We first perform the truncation on a CPU, as handling conditional statements is not a strong point for GPU.  The truncated 
 array is then send to the GPU for Hough transform.  The GPU can perform this computation by distributing it over the 
 large number of cores available.  Suppose after  truncation the number of remaining pixels is $N$, 
 with $N  \le N_t N_f$, and number of DM trial value $N_d$, we may launch for example a total of $N N_d$ threads, 
 each compute the accumulator values for this pixel. In this case the GPU computing cores  and the threads need to 
 share the common memory of the total accumulator array, which is a sum of the accumulator array of all threads, 
  but otherwise the computation of the threads are independent.

\begin{figure}
 \centering
  \includegraphics[trim=10 10 40 34,clip,width=0.4\textwidth]{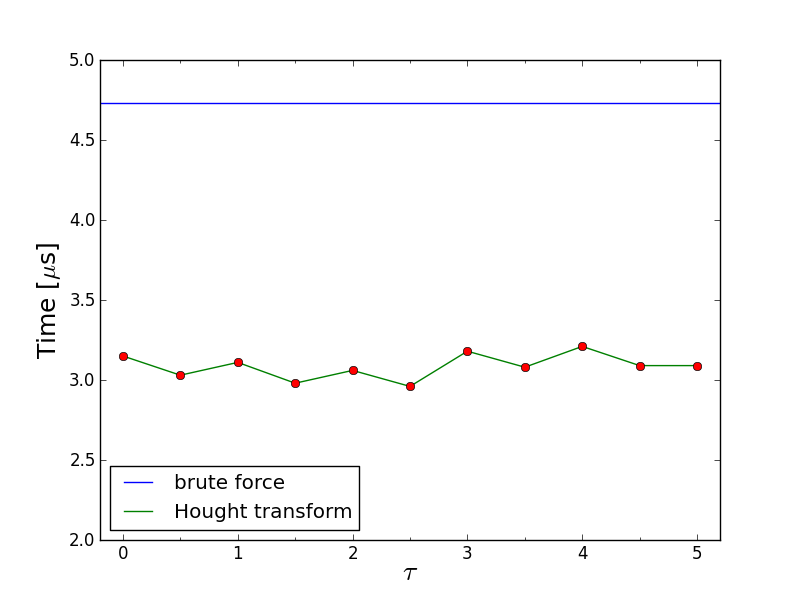}
 \caption{The computation time with GPU as a function of truncation threshold $\tau$.}
  \label{fig:GPUtime}
\end{figure}

In Fig.~\ref{fig:GPUtime} we show the computation time using a GPU server. The GPU we used is an NVIDIA GTX 1080TI. 
For the test the data frame has a size of $N_f=1024, N_t=512, N_d=2000$. We plot both the time needed for the brute force 
method and the Hough transform method. As can be seen from the figure, compared with the CPU, the GPU computation 
takes very short time. The Hough transform takes slightly more than half of the time of the brute force computation, and unlike
the CPU case,  it is almost independent of the truncation threshold. This is because in the case of GPU, although $N$ is larger 
for lower threshold $\tau$, for this parameter range the GPU can launch a total number of $N N_d$ threads.
The time needed for computation is essentially the time needed for one thread to complete its computation, 
plus the time for each process to communicate its result, which is a very small amount in the present case. 
Thus, with GPU one can achieve high sensitivity by choosing a low truncation threshold.  
However, as point out above, for large arrays many beams may be formed and feed to the GPU for computation. 
We should also note that In a practical implementation,  there will also be many restrictions which 
reduce the computing speed, for example the data transfer rate may severely 
limit the amount of the data that a GPU can handle, so the computation speed is much slower than indicated 
by Fig.\ref{fig:GPUtime}.  Reducing the computational complexity is still quite  meaningful at present.

In this paper we have focused primarily on the concept and theoretical
  aspects of the Hough transform method for radio burst detection.
  We demonstrated the application of this method by
  simulation and some real pulsar and FRB data, though for real application
there are still some practical issues to be addressed, for example,
determining the implementation and optimisation of the Hough transform
algorithm, the parallelism and acceleration strategy; selecting the
appropriate truncation threshold as a  trade-off
between the sensitivity and the computation time; developing
peak searching algorithm which can integrate the neighbouring pixels with different kernel size, etc.
These are beyond the scope of this paper, we leave them for future studies.

\section*{Acknowledgements}
We thank Fengquan Wu and Chenhui Niu for discussions.
This work is supported by the National Natural Science Foundation of China (NSFC) grant 11633004 and the NSFC-ISF(Isreal 
Science Foundation) grant 11761141012, the Ministry of Science and Technology (MoST) grant 2016YFE0100300 and 
2018YFE0120800,  the Chinese Academy of Science (CAS) Frontier Science Key Project QYZDJ-SSW-SLH017,
and the National Astronomical Observatory of China (NAOC) pilot research grant Y834071V01.




\bibliographystyle{mnras}
\bibliography{hough} 


\bsp	
\label{lastpage}
\end{document}